\documentclass[12pt]{elsarticle}

\journal{Phys. Lett. B}

\usepackage{graphicx}
\newcommand{\MeV}{\mbox{MeV}}
\newcommand{\GeV}{\mbox{GeV}}

\begin{document}

\begin{frontmatter}

\title{A lattice QCD calculation
of the transverse decay constant of the $b_1(1235)$ meson.
}

\author[ZEUT]{K. Jansen}
\author[GLA]{C. McNeile}
\author[LIV]{C. Michael}
\author[BONN]{C. Urbach}

\address[ZEUT]{
NIC, DESY, Zeuthen, Platanenallee 6, D-15738 Zeuthen, Germany
}

\address[GLA]{
Department of Theoretical Physics, University of Wuppertal,
Wuppertal, 42199, Germany.
}

\address[LIV]{
Theoretical Physics Division, Dept. of Mathematical Sciences,
University of Liverpool, Liverpool L69 3BX, UK
}

\address[BONN]{
Humboldt-Universit\"{a}t zu Berlin,  Institut f\"{u}r Physik
Mathematisch-Naturwissenschaftliche Fakult\"{a}t I,
Theorie der Elementarteilchen / Ph\"{a}nomenologie,
Newtonstr. 15, 12489 Berlin Germany
}

\begin{abstract}
We review various B meson decays that require knowledge
of the transverse decay constant of the $b_1(1235)$ meson.
We report on an exploratory
 lattice QCD calculation of 
the transverse decay constant of the $b_1$ meson.
The lattice QCD calculations used unquenched
gauge configurations, at two lattice spacings, 
generated with two flavours of sea
quarks. The twisted mass formalism is used.
\end{abstract}

\begin{keyword}



\end{keyword}

\end{frontmatter}

\section{Introduction and motivation}

The transverse decay constant of the $b_1(1235)$ meson 
($f_{b_1}^T$)
is theoretical
input to a number of decays of the B meson.
For example $f_{b_1}^T$ 
is an important QCD input to the following decays:
$\overline{B}^0 \rightarrow b_1^- \rho^+$, 
$\overline{B}^0 \rightarrow b_1^- K^{\star +}$,
using the light cone formalism~\cite{Yang:2005tv,Cheng:2007mx}.
The $f_{b_1}^T$ constant is also input to the
decay $B \rightarrow b_1 \gamma$~\cite{JamilAslam:2006bw,Wang:2007an}
that use light cone sum rules.
Diehl and Hiller~\cite{Diehl:2001xe} discuss studying decays
of the B meson with final states that include the
$b_1$ meson. 

There are alternative 
theoretical 
formalisms~\cite{Laporta:2006uf,Calderon:2007nw,Wang:2008hu,Munoz:2009ii}
to the light cone sum rules
such as factorization,
that describe the non-leptonic decays 
of the B meson to final states that 
include the $b_1$ meson. So it is important to have 
a cross-check of the input parameters used in 
the light cone formalism.

BaBar has experimentally measured the B decays:
$b_{1}\pi$ and $b_1 K$~\cite{Aubert:2007xd,Aubert:2008eq}.
The charmless decays of the B meson,
that include those with a $b_1$ meson in the final state,
 have been reviewed
by Cheng and Smith~\cite{Cheng:2009xz}.

The transverse decay constant of the $b_1$ meson is not accessible
to experiment, but can be calculated in models~\cite{Chizhov:2003qy}
and sum rules. Calculations
of the $b_1$ meson have also been used to tune
sum rules~\cite{Bakulev:1999gf,Broniowski:1998ws,Govaerts:1986ua}.
In particular, the same sum rules are used to 
simultaneously extract the transverse decay constants
of the $b_1$ and 
$\rho$ mesons~\cite{Bakulev:1999gf,Broniowski:1998ws}.

In principle lattice QCD should be able to produce an
accurate result for $f_{b_1}^T$, particularly
as modern lattice QCD calculations usually have
multiple lattice spacings and volumes, with pion masses
below 300 \MeV~\cite{Jansen:2008vs}.
To the best of our knowledge,
there has never been a lattice QCD calculation
of $f_{b_1}^T$ before this one.

The $b_1$ meson is a good test case for lattice techniques
that deal with resonances, because it is thought to be 
a basic quark-antiquark meson that decays via S-wave.
The experimental width of the $b_1$(1235) is 142(9) \MeV,
and the bulk of the decays are to $\omega \pi$.
Hence further  motivation for this study is to compute as much
information about the $b_1$ meson from our lattice
QCD calculations as possible.

Light cone sum rules and factorization methods,  also use the 
decay constants of the $a_0$, $\pi(1300)$,
$a_1$ mesons to study the decays of 
the B meson, but there have been previous lattice QCD
calculations of those 
quantities~\cite{Wingate:1995hy,McNeile:2006nv,McNeile:2006qy}.

\section{The lattice QCD calculation}

The transverse decay constant ($f_{b_1}^T(\mu)$) of the $b_1$ 
meson is defined~\cite{Yang:2007zt}
by
 \begin{equation}
\langle 0 \mid 
\overline{\psi} \sigma_{\mu \nu} \psi
\mid b_1 (P,\lambda) \rangle = 
i f_{b_1}^T(\mu) \epsilon_{\mu \nu \alpha \beta} \epsilon^{\alpha}_{(\lambda)}
P^\beta
\label{eq:TRANSrhodecayDEFN}
 \end{equation} 
 where $\sigma_{\mu \nu} = i/2 [\gamma_\mu , \gamma_\nu] $,
and $\epsilon^{\alpha}_{(\lambda)}$ is the polarisation vector of
the meson.
It is
convenient to introduce the tensor  current $T_{\nu \mu} =
\overline{\psi} \sigma_{\mu \nu} \psi$. 
We do not include any momentum in this lattice calculation.
For completeness we note that the $b_1$ meson has $J^{PC} = 1^{+-}$.

In the isospin limit the leptonic decay constant 
of the $b_1$ meson is zero, because the $T_{ij}$ 
operator is orthogonal to the vector and axial currents. 
Although for some quantities, such as the matrix element for $\rho-\omega$
mixing~\cite{McNeile:2009mx}
or the decay constant of the flavour non-singlet
$0^{++}$ meson~\cite{McNeile:2006nv}
an estimate of isospin
violating quantities can be made 
from a lattice calculation with two degenerate
flavours of sea quarks, we don't see how
to estimate the leptonic decay constant of the
$b_1$ meson without using non-degenerate light
quarks. A formalism to do this has recently 
been developed for twisted mass QCD~\cite{WalkerLoud:2009nf}.

Our lattice calculation uses the twisted  mass QCD
formalism~\cite{Frezzotti:2000nk}. Once a single parameter has been
tuned, twisted mass QCD has non-perturbative $O(a)$ 
improvement~\cite{Frezzotti:2003ni,Dimopoulos:2009qv}. 
The twisted mass formalism has been reviewed by
Shindler~\cite{Shindler:2007vp}.

\begin{table}[tb]
\centering
\begin{tabular}{|c|c|c|c|c|} \hline
Ensemble &  $\beta$  & $a \mu_q$ & Volume & $f^T_{b_1}(2 \GeV) $ \\ \hline
$B_1$ &  3.9  & 0.004 &  $24^3 \times 48$   & 249(55) \\
$B_2$ &  3.9  & 0.0064 &  $24^3 \times 48$  & 239(29) \\
$B_3$ &  3.9  & 0.0085 &  $24^3 \times 48$  & 254(26) \\
$B_4$ &  3.9  & 0.0100 &  $24^3 \times 48$  & 220(32) \\
$B_5$ &  3.9  & 0.0150 &  $24^3 \times 48$  & 256(33) \\
$B_6$ &  3.9  & 0.004  & $32^3 \times 64$  & 233(29) \\
$C_1$ &  4.05  & 0.003 & $32^3 \times 64$  & 202(83)  \\
$C_2$ &  4.05  & 0.006 & $32^3 \times 64$  & 193(55) \\
$C_3$ &  4.05  & 0.008 & $32^3 \times 64$  & 216(44) \\
$C_4$ &  4.05  & 0.012 & $32^3 \times 64$  & 289(47) \\
\hline
\end{tabular}
\label{tb:fb1results}
  \caption{
Summary of results for $f_{b_1}^T$ used in this calculation.
$a \mu_q$ is the bare mass of the light quark in
lattice units. The ensemble names are from~\cite{Urbach:2007rt}.
}
\end{table}

We have recently reported on the some basic measurements
of the $\rho$, $b_1$ and $a_0$ mesons~\cite{Jansen:2009hr}. 
In this paper we extend that study to the transverse
decay constant of the $b_1$ meson. All the necessary
details are in the previous paper~\cite{Jansen:2009hr}
and here we provide a brief summary. There is a additional
information about the lattice techniques, such as the smearing
and variational analysis in the "methods paper for the ETM 
collaboration~\cite{Boucaud:2008xu}." 
We used the twisted mass Wilson action and the tree level
improved Symanzik action.
The ensembles 
used in this analysis are in table~\ref{tb:fb1results}.
Details of the analysis of light pseudo-scalar 
mesons are in~\cite{Boucaud:2008xu,Blossier:2007vv}.

The correlators used to extract the $f_{b_1}^T$
decay constant are in equation~\ref{eq:TT}. We use
a smearing matrix of order 2, using basis functions
of local and fuzzed interpolating operators,
that includes the correlators in equation~\ref{eq:TT}.
We fit the smearing matrix to a factorizing 
fit form~\cite{Boucaud:2008xu} with two states.
The charged interpolating operator for the $b_1$
meson in the twisted basis was used~\cite{Boucaud:2008xu}.

\begin{equation}
\sum_x \sum_{k=1, i < j}^{3} \epsilon_{ijk}
\langle T_{ij}(x,t_x) T_{ij}(0,0)^\dagger \rangle
\rightarrow
\frac{3 m_{b_1} (f_{b_1}^T)^2 e^{-m_{b_1} t_x}  }{2 }
\label{eq:TT}
\end{equation}

\begin{figure}
\begin{center}
\includegraphics[scale=0.5,angle=0]{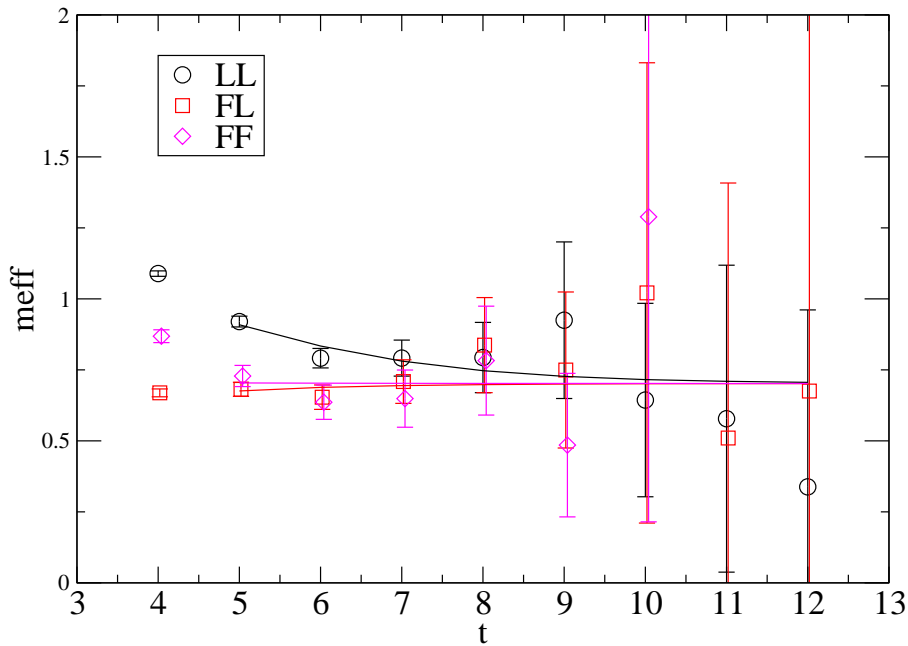}
\end{center}
\caption {
Effective mass plot for the $b_1$ channel
at $\beta$ = 3.9, $\mu_q=$0.004, L=24.  The labels LL, FL, FF 
refer to local-local, fuzzed-local, and fuzzed-fuzzed.
}
\label{fig:meff}
\end{figure}

\begin{figure}
\begin{center}
\includegraphics[scale=0.5,angle=0]{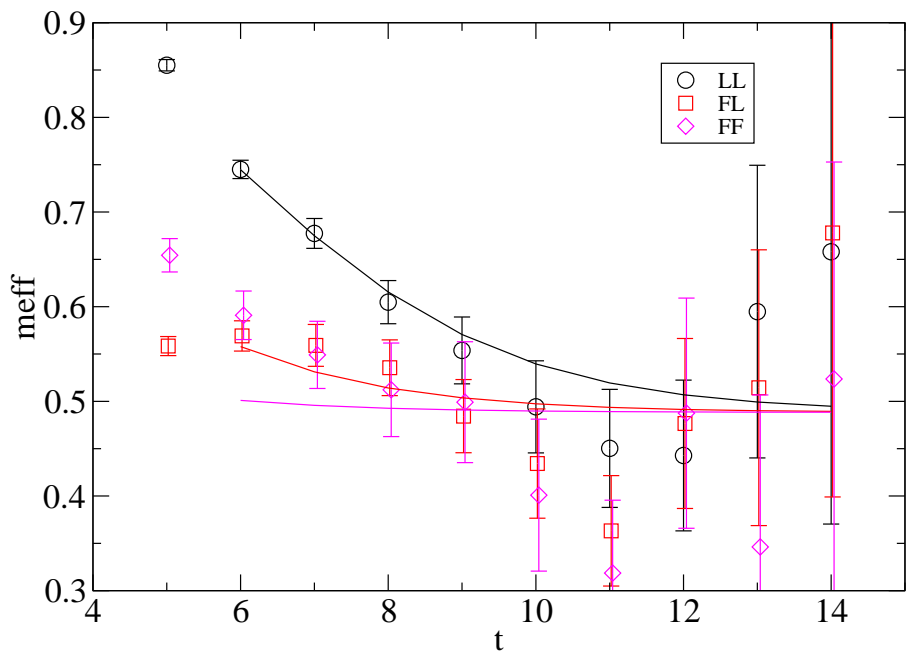}
\end{center}
\caption {
Effective mass plot for the $b_1$ channel
at $\beta$ = 4.05, $\mu_q=$0.006, L=32.  The labels are the
same as for figure~\ref{fig:meff}.
}
\label{fig:mefffine}
\end{figure}

We used ensembles at two different
$\beta$ values. At $\beta=3.9$ we included two
volumes.
We used the pion decay constant of the $\pi$
meson to determine the lattice spacing. 
At $\beta=4.05$ ($\beta=3.9$) the lattice spacing
is: a = 0.0667(5) fm 
(a = 0.0855(5) fm). The lattice spacing from $f_\pi$ was
consistent with the value from the nucleon mass~\cite{Alexandrou:2008tn}.
In figures~\ref{fig:meff} and ~\ref{fig:mefffine} we show  effective mass plots
for the $b_1$ correlators.

The decay constant $f_{b_1}^T$ depends on the value
of the renormalisation scale. We used a renormalisation
factor obtained from the Rome-Southampton
non-perturbative 
method~\cite{Martinelli:1994ty,Dimopoulos:2007fn,Dimopoulos:2009prep}.
As traditional in lattice 
QCD calculations we quote the result at the scale of 2 GeV.
The renormalisation group equations can be used to evolve 
the decay constant to another scale.

\begin{figure}
\begin{center}
\includegraphics[scale=0.5,angle=0]{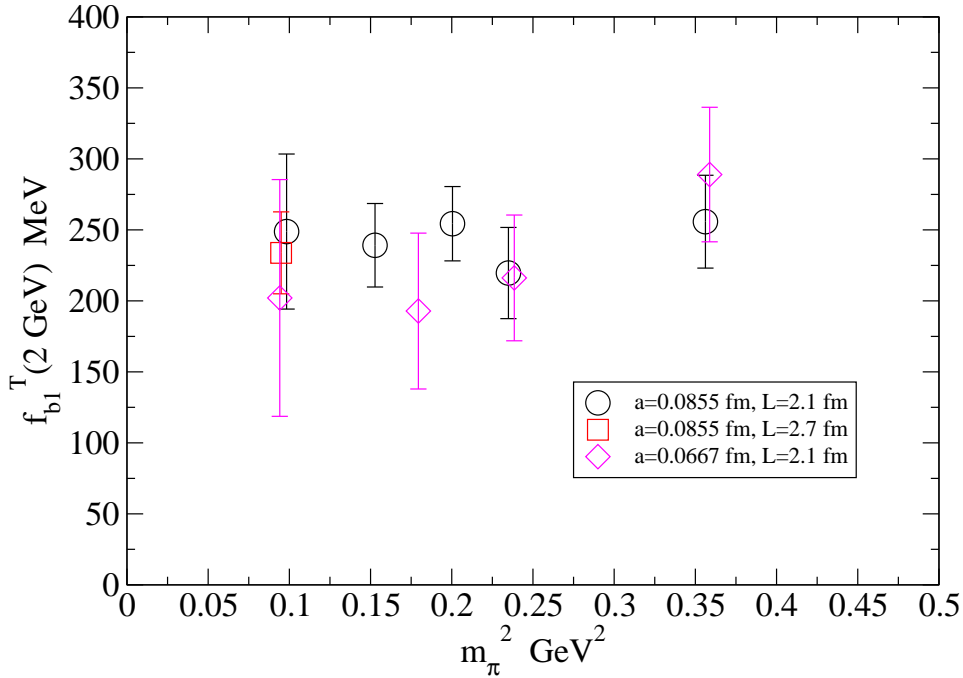}
\end{center}
\caption {
The decay constant of the $b_1$ meson as a function of the
square of the pion mass.
}
\label{fig:f1decay}
\end{figure}

The results for the decay constant from this calculation
are in table~\ref{tb:fb1results}.
In figure~\ref{fig:f1decay} the
transverse decay constant
of the $b_1$ meson is plotted  in physical units as a function of the 
square of the pion mass. Although the error bars are large,
the results for $f_{b_1}^T$ are consistent between the two
lattice spacings and volumes.

A common technique to check whether a state is a
scattering state or a resonance is look at the 
volume dependence of the amplitude~\cite{Mathur:2004jr}.
The use of the volume dependence of the amplitude
and the connection to L\"{u}scher's method has recently been
discussed by Meng et al.~\cite{Meng:2008zzb}. 
The "rule of thumb" is that if our
$b_1$ correlator couples to a scattering state
of $\omega \pi$, the volume dependence of
the $f_{b_1}^T$ decay constant extracted from
equation~\ref{eq:TT} is
\begin{equation}
f_{b_1}^T \sim \frac{1}{\sqrt{V}}
\label{eq:resvol}
\end{equation}
where $V$ is the spatial volume.
For a resonance $f_{b_1}^T$ should be independent
of the volume (apart from small corrections if the 
box size is too small to fit the resonance state).

The numerical results at $\mu_q=0.004$ at
$\beta=3.9$, show that $f_{b_1}^T(L=32)/f_{b_1}^T(L=24)$
= 0.94(24), compared to the prediction for scattering
states in equation~\ref{eq:resvol} of 0.65.

In our previous paper~\cite{Jansen:2009hr} we showed that the 
decay of the $b_1$ meson to $\omega \pi$ was open in our calculation.
However the mass of the lightest state in the $1^{-+}$
channel didn't track the sum of the masses of the
$\omega$ and $\pi$ mesons particularly well.
In L\"{u}scher's formalism for the
study of resonances on the lattice the opening
of strong decays is described by an avoided level
crossing~\cite{Luscher:1991cf}.
The detailed calculations of Bernard 
et al.~\cite{Bernard:2008ax} for the $\Delta$ baryon
suggested that the avoided level crossing is "washed out"
by the dynamics, so comparing the mass from the resonant interpolating
operator to the sum of the masses of the decay products 
is probably too simplistic. A similar situation happened with
string breaking, where the linearly rising potential 
was seen to increase beyond the energy that allowed the
string to break to two static B mesons~\cite{Bali:2005fu}.

The above considerations suggest that although we don't have
full control over the resonant nature of the $b_1$ meson,
our interpolating operators are coupling to the $b_1$ meson
in the range of quark masses in our calculations.
We extrapolate $f_{b_1}^T$ linearly in the square of the 
pion mass to get $f_{b_1}^T(2~\GeV)$ = 236(23) \MeV~at the physical 
pion mass at $\beta=3.9$.

\begin{table}[tb]
\centering
\begin{tabular}{|c|c|c|} \hline
Group & Method &  $f_{b_1}^T$(1 \GeV) MeV  \\ \hline
Chizhov~\cite{Chizhov:2003qy} & extended NJL  quark model & 175(9) \\
Ball and Braun ~\cite{Ball:1996tb} & sum rule & 180(20) \\
Bakulev and Mikhailov~\cite{Bakulev:2000er,Bakulev:1998pf} & sum rule & 184(5) \\
Bakulev and Mikhailov~\cite{Bakulev:2000er,Bakulev:1998pf} & sum rule & 181(5) \\
Yang~\cite{Yang:2005gk,Yang:2007zt} & sum rule & 180(8) \\
\hline  \hline
This calculation & lattice QCD & 258(25) \\
\hline
\end{tabular}
\label{tb:fb1summary}
  \caption{
 Summary of calculations of $f_{b_1}^T$ at 1 \GeV.
}
\end{table}

In table~\ref{tb:fb1summary} we collect together other
estimates for $f_{b_1}^T$.
We have evolved our lattice result to the scale of
1 \GeV~ to compare with the results from sum rules.
The formalism to evolve the decay constant with scale
is described~\cite{Jansen:2009hr},
that uses input from perturbative calculations by
Gracey and 
others~\cite{Gracey:2000am,Chetyrkin:2000yt,vanRitbergen:1997va,Czakon:2004bu}.
The perturbative factor is 1.095 to evolve $f_{b_1}^T$
from 2 \GeV~ to 1 \GeV.

In the non-relativistic quark model the $b_1$ meson is a P-wave
meson with a node in the wave-function at the origin. This would
suggest the transverse decay constant is small. 
The partial inclusion of relativistic 
effects in the quark model~\cite{Hayne:1981zy}
increases the decay constant.
The results
in table~\ref{tb:fb1summary} show that $f_{b_1}^T$ is of the same
order of magnitude as the pion decay constant (132 MeV), so this is
evidence that local interpolating operators will couple
to the $b_1$ meson. Pragmatically using 
derivative sources~\cite{Gattringer:2008be,Engel:2009cq}
with smearing techniques, such as Jacobi, may be 
useful to get a good signal.

\section{Conclusion}

We have presented the first calculation of the transverse
decay constant of the $b_1$ meson from lattice QCD. 
We obtain
$f_{b_1}^T(2~\GeV)$ = 236(23) MeV at the physical 
pion mass.
The
result is higher than the sum rule results by 3 $\sigma$.
Future lattice QCD calculations need to
reduce the statistical errors on the correlators, 
and to directly 
take into account the resonant nature of the $b_1$ meson.

\section{Acknowledgments}

We thank all members of ETMC for a very fruitful collaboration.
We thank Prof. Braun for providing the error for the result 
in~\cite{Ball:1996tb}, and we thank Mihail Chizhov for useful
comments on the paper.


\end{document}